\documentclass[aip,jcp,amsmath,reprint]{revtex4-1}
\usepackage{natmove}

\usepackage{graphicx}
\usepackage[update,prepend]{epstopdf}
\newcommand{\vv}[1]{\mathbf{#1}}

\begin{document}
\title{Stratification of polymer mixtures in drying droplets: hydrodynamics and diffusion}
\author{Michael P. Howard}
\email{mphoward@utexas.edu}
\affiliation{McKetta Department of Chemical Engineering, University of Texas at Austin, Austin, TX 78712, United States}

\author{Arash Nikoubashman}
\email{anikouba@uni-mainz.de}
\affiliation{Institute of Physics, Johannes Gutenberg University Mainz, Staudingerweg 7, 55128 Mainz, Germany}

\begin{abstract}
We study the evaporation-induced stratification of a mixture of short and long
polymer chains in a drying droplet using molecular simulations.
We systematically investigate the effects of hydrodynamic interactions (HI)
on this process by comparing hybrid simulations accounting for HI between polymers through the
multiparticle collision dynamics technique with free-draining Langevin dynamics
simulations neglecting the same. We find that the dried supraparticle morphologies
are homogeneous when HI are included but are stratified in
core--shell structures (with the short polymers forming the shell) when HI are neglected.
The simulation methodology unambiguously attributes this difference
to the treatment of the solvent in the two models. We rationalize the
presence (or absence) of stratification by measuring phenomenological multicomponent
diffusion coefficients for the polymer mixtures. The diffusion coefficients
show the importance of not only solvent backflow but also HI between polymers
in controlling the dried supraparticle morphology.
\end{abstract}

\maketitle
\section{Introduction}
\label{sec:intro}
Understanding microstructure formation during the drying of a volatile solvent
from a mixture is essential for engineering coatings \cite{Routh:2013dm}, polymer
nanocomposites \cite{Kumar:2017wz}, and nanocrystal superlattices \cite{Boles:2016et}, among many technologies.
The dried microstructure depends on multiple controllable parameters, including
processing conditions like temperature \cite{Ye:2001dl,Im:2002fp}, relative humidity \cite{Chung:2006jr},
and solvent properties \cite{Courty:2011hr}. For example, when these conditions cause fast drying,
colloidal suspensions densify and crystallize from the solvent--air interface,
while under slow drying conditions, the crystal nucleates from the bulk
\cite{Narayanan:2004cr,Bigioni:2006hf,Wang:2017vs,Howard:2018fb}; the quality of the
resulting crystal depends on the specific processing pathway \cite{Im:2002fp,Cheng:2013wt}.
Multicomponent mixtures exhibit even richer behavior because their constituents can
compositionally segregate during drying \cite{Schulz:2018ff}, holding great promise for
assembling functional materials through simple single-step processing.

In particular, the formation of layered films from mixtures of differently sized solutes
has received considerable recent attention \cite{Schulz:2018ff}. Film morphology plays an important
role in, e.g., tuning the refractive and reflective characteristics of optical
materials,\cite{chhajed:apl:2008, lee:ami:2016} improving the properties
of pressure-sensitive adhesives,\cite{carelli:adh:2007} and fabricating superhydrophobic
coatings.\cite{lopez:poly:2017} Experiments and computer simulations showed
that well stabilized mixtures of colloidal particles\cite{Trueman:2012ve,Fortini:2016ip,%
MartinFabiani:2016fj,Makepeace:2017ht,Liu:2018vb,Carr:2018gl,Liu:2019fq,Howard:2017bq,%
Fortini:2017jl,Tang:2018do,Tang:2019kh,Tang:2019jy},
polymers \cite{Howard:2017vu,Statt:2018bw}, or both \cite{Howard:2017vu,Cheng:2016}.
will stratify by size during fast drying to form films with a top layer enriched
in the smaller component. Similar ``small-on-top'' stratification also occurs in
evaporating droplets, which can be exploited to produce supraparticles with core--shell
morphologies.\cite{raju:langmuir:2018, liu:acsnano:2019, gartner:msde:2020}
These supraparticles are promising for applications in catalysis due to their distinctive
hierarchical structures\cite{hou:aml:2020} and tunable porosity\cite{liu:acsnano:2019b}
or as structural colorants \cite{Xiao:2019uo}.

In order to understand and control the morphology of such materials,
different theoretical models were proposed to predict microstructure and stratification
during drying \cite{Routh:2004jz,Trueman:2012er,Fortini:2016ip,Howard:2017bq,Zhou:2017gg}.
Previously, we developed one model \cite{Howard:2017bq,Howard:2017vu} using the framework of dynamic
density functional theory (DFT) \cite{Marconi:1999wb,Archer:2004eq,Archer:2005wv}. This model belongs to a
general class of multicomponent diffusion models for stratification \cite{Trueman:2012er,Zhou:2017gg}.
In experiments, temperature and pressure gradients inside the drying mixture typically
relax faster than the solutes diffuse, so the temperature and pressure can be
considered approximately constant. The theories then postulate that
the diffusive flux $\vv{j}_i$ of each component $i$ in an $n+1$ component system
(comprising $n$ solutes and the solvent) can be generally written as
\begin{equation}
	\vv{j}_i = \rho_i (\vv{u}_i - \bar{\vv{u}}) = -\sum_{j=0}^n L_{ij} \nabla \mu_j,
	\label{eq:onsager}
\end{equation}
where $\rho_i$ and $\vv{u}_i$ are the number density and average velocity of component $i$,
respectively, $\bar{\vv{u}}$ is the velocity of a chosen reference frame, $L_{ij}$ is a
phenomenological Onsager coefficient, and $\mu_j$ is the chemical potential of component $j$.
The Onsager coefficients must be measured (or assumed) and are symmetric ($L_{ij} = L_{ji}$)
if Onsager's reciprocal relations hold \cite{Onsager:1931uu}. The chemical potential gradients are thermodynamic
driving forces for diffusion that are self-generated by the drying process and can be
computed based on the solute concentration profiles using, e.g., virial expansions of the
free energy \cite{Zhou:2017gg} or more accurate free-energy functionals \cite{Howard:2017bq,Howard:2017vu}.

The reference frame can be simply chosen as the stationary laboratory ($\bar{\vv{u}} = \vv{0}$),
but it is often convenient to define the fluxes relative to a velocity
that is a weighted function of $\vv{u}_i$ \cite{Brady:1975vd}. The volume-averaged velocity
$\bar{\vv{u}}_{\rm v} = \sum_{i=0}^n v_i \rho_i \vv{u}_i$ (with $v_i$ being the volume of
component $i$) is particularly expedient if the drying mixture is
incompressible because $\bar{\vv{u}}_{\rm v} = \vv{0}$ so that there is no net
volume flux. Making use of the reciprocal relations and that
$\sum_{i=0}^n v_i \vv{j}_i = \vv{0}$ in this reference frame yields
\begin{equation}
	\vv{j}_i = -\sum_{j=1}^n L_{ij} \nabla \mu_j',
	\label{eq:exchonsager}
\end{equation}
where $\mu_j' = \mu_j - (v_j/v_0)\mu_0$ is an \textit{exchange} chemical
potential that gives the free-energy change to insert a solute $j$ and remove an
equivalent volume of solvent, designated as component 0. However, the chemical potential gradients are
not generally independent of each other because they satisfy the Gibbs--Duhem
relationship, $\sum_{i=0}^n \rho_i \nabla \mu_i = 0$ at constant temperature and pressure,
in local thermodynamic equilibrium.
The solvent can then be eliminated from Eq.~\eqref{eq:onsager} by redefining
\begin{equation}
	\vv{j}_i = -\sum_{j=1}^n \Lambda_{ij} \nabla \mu_j,
	\label{eq:effonsager}
\end{equation}
where we stress that the solvent chemical potential $\mu_0$ and flux $\vv{j}_0$
are no longer considered independent, and new effective
coefficients $\Lambda_{ij}$ have been defined from the original $L_{ij}$.
Equation~\eqref{eq:effonsager} can be derived in other reference frames, even
in one where all $\vv{j}_i$ are independent and Eq.~\eqref{eq:exchonsager} does
not hold, so we will focus on this effective formulation.

Within such multicomponent diffusion models, small-on-top stratification occurs when
the total driving force on the larger solute
is sufficiently strong compared to that on the smaller solute so that the two separate as
they codiffuse \cite{Fortini:2016ip}. We and others \cite{Howard:2017bq,Howard:2017vu,Zhou:2017gg} effectively assumed dominant diagonal couplings,
$\Lambda_{ii} \approx \rho_i D_i/(k_{\rm B}T)$, with negligible off-diagonal couplings,
$\Lambda_{i,j\ne i} \approx 0$, where $k_{\rm B}$ is Boltzmann's constant, $T$ is the
temperature, and $D_i$ is the equilibrium self-diffusion coefficient of component $i$.
When cross-interactions between solutes were included
in the chemical potentials \cite{Zhou:2017gg}, these models predicted stratification qualitatively
resembling that observed in experiments \cite{Fortini:2016ip,Makepeace:2017ht,Liu:2018vb} and were even in quantitative
agreement with free-draining implicit-solvent computer simulations of drying
polymer mixtures \cite{Howard:2017vu}. However, it was recently demonstrated that these models
overpredict the extent of stratification compared to experiments of drying mixtures
of colloidal particles \cite{Schulz:2018ff,liu:acsnano:2019}.
The overprediction is thought to originate from solvent effects
like backflow and/or hydrodynamic coupling that are missing from the models \cite{Brady:2011bh,Sear:2017cv}.

To remedy this shortcoming, an alternative explanation for stratification was
proposed based on the concept of diffusiophoresis \cite{Sear:2017cv,Sear:2018dn}.
Sear and Warren analyzed the migration of a single, infinitely large colloid in
an ideal polymer solution with hard excluded-volume interactions between the
colloid and polymers \cite{Sear:2017cv}.
The solvent was taken into account using a continuum description of a thin
film flow near the surface of the colloid that is excluded to the
polymers. In this picture, the colloid migrates towards regions of lower polymer
density with diffusiophoretic velocity
\begin{equation}
	\vv{u}_{\rm c} = - \frac{R_{\rm g}^2 k_{\rm B} T}{2 \eta} \nabla \rho_{\rm p},
	\label{eq:dp}
\end{equation}
where $R_{\rm g}$ is the radius of gyration of the polymer, $\eta$ is the viscosity,
and the subscripts ``c'' and ``p'' denote the colloid and polymers, respectively.
Stratification occurs when the large colloid moves down the polymer gradient
faster than the polymers themselves.
The diffusiophoretic velocity of the colloid can be rewritten as a flux in the form
of Eq.~\eqref{eq:effonsager},
\begin{equation}
	\vv{j}_{\rm c} = -\frac{R_{\rm g}^2\rho_{\rm c}}{2\eta}\nabla \rho_{\rm p}.
\end{equation}
This flux might be identified as an off-diagonal coupling,
$\Lambda_{\rm cp} = R_{\rm g}^2 \rho_{\rm c}\rho_{\rm p}/(2\eta)$,
between the chemical potential gradient of the polymer and the flux of the colloid
by assuming $\Lambda_{\rm cc} \approx 0$ and that the polymer chemical
potential is dominated by its own (ideal) contributions,
$\nabla \mu_{\rm p} \approx k_{\rm B} T \nabla \ln (\lambda_{\rm p}^3 \rho_{\rm p})$
where $\lambda_{\rm p}$ is the thermal wavelength of the polymer.
We note, however, that this splitting is not unique because $\nabla\rho_{\rm p}$ contributes
to both $\nabla \mu_{\rm c}$ for the colloid and $\nabla \mu_{\rm p}$ for the polymers
in nonideal mixtures, and there may be additional terms in $\Lambda_{\rm cc}$ and
$\Lambda_{\rm cp}$ that effectively cancel each other \cite{Sear:2017cv}.

Predictions based on the diffusiophoresis model are in better agreement with
experiments than prior diffusion models but are still not quantitatively accurate \cite{Schulz:2018ff}.
The reason for this discrepancy is not obvious, but may be due to some of the
approximations that were required to render the calculation of $\vv{u}_{\rm c}$
analytically tractable. For example, solutes have finite size ratios in experiments (typically
10:1 or smaller), but the model assumes that the colloid is much larger than the polymers.
Larger solutes like the colloid are also not infinitely dilute in experiments,
and so can interact with each other in ways that are not easily accounted for
within this model. Generally, most of the mixtures that have been studied experimentally are not sufficiently
dilute that they can be assumed to be thermodynamically ideal as in the Asakura--Oosawa
treatment of the polymers \cite{Sear:2017cv}, particularly for colloidal mixtures and/or once the mixture concentrates during drying.
To date, nonidealities like skin-layer formation \cite{Okuzono:2006gf} or jamming \cite{Sear:2018dn} have only
been taken into account in an ad hoc fashion. Last, the diffusiophoretic picture
becomes more complex for solutes that can be penetrated by solvent and/or deform
(e.g, polymer mixtures), as this complicates the analysis of the fluid flow \cite{RamirezHinestrosa:2020gd}.

Computer simulations taking into account solvent effects and
hydrodynamic interactions (HI) can play a key role in addressing some of these
questions \cite{howard:coce:2019}. Simulations can model the solute, size, or concentration regimes
that are highly relevant to experiments but are not amenable
to a purely theoretical analysis. Simulations also resolve microscopic detail
that can be used to stringently test theoretical models and identify key physics,
both thermodynamic and hydrodynamic in nature,
that are required to improve them. For example, chemical potentials, which are
challenging to measure in experiments, can be computed directly in simulations
using, e.g., Widom's test insertion method \cite{Widom:1963gu}.
Different treatments of solvent-mediated interactions can also be systematically
included or excluded from a simulation model in ways that cannot be achieved
in experiments. This approach has been used to show the importance of HI in setting the
microstructure of drying colloidal suspensions \cite{Howard:2018fb,Chun:2020ic}, but
their role in stratification is still debated \cite{Tang:2019jy}.

One of us previously used both explicit-solvent and implicit-solvent molecular
simulations to probe the role of HI in the stratification of drying polymer mixtures
in a thin film \cite{Statt:2018bw}. The polymers were initially dissolved in a solvent
explicitly represented as a Lennard-Jones fluid in vapor--liquid coexistence.
The explicit solvent not only propagated HI, but also contributed to the effective
interactions between the polymers. The implicit-solvent simulations neglected
hydrodynamic coupling between polymers, but importantly, both the effective
interactions between polymers and their equilibrium self-diffusion coefficients
were matched at infinite dilution. It was found, after extensive testing of the
two models, that the presence of HI effectively
suppressed small-on-top stratification. Although ultimately effective, one
potential challenge of comparing separate explicit-solvent and implicit-solvent
simulation models like these is that thermodynamic and hydrodynamic effects can
become convoluted if the model interactions
are not perfectly matched and/or if the implicit-solvent model has limited
transferability to conditions at which it was not parameterized. Interfacial
effects were also handled differently between the two models \cite{Tang:2018ff},
leaving open questions about their role in setting the microstructure.
Therefore, it is preferable to make comparisons like these between simulations using exactly
the \textit{same} effective interactions for the solutes and interfaces so that the only
difference between models is the presence or absence of HI.

Given this context, we set two goals for this work. Our first aim was to explore
the possibility of stratification of a mixture of short and long polymers in a
drying droplet, which has not been previously investigated. We simulated this
process for the same coarse-grained polymer model using two different
treatments of the solvent and corresponding polymer dynamics: one incorporating HI
between polymers and one neglecting the same (Sec.~\ref{sec:model}).
Consistent with prior work \cite{Statt:2018bw}, the mixture stratified to form a core--shell supraparticle
morphology when HI were neglected, but did not stratify when
HI were included (Sec.~\ref{sec:evap}). Our second aim was to rationalize this
behavior using a multicomponent diffusion model. We used nonequilibrium simulations
to measure the transport coefficients that couple the polymer fluxes to their
chemical potential gradients (Sec.~\ref{sec:onsager}), finding qualitatively
different behaviors between the two solvent treatments
that were consistent with the presence or absence of stratification. Our study
strongly supports the importance of HI in stratification phenomena, and we advocate
incorporating these interactions in future theoretical and computational models.

\section{Model and Methods}
\label{sec:model}
The polymers comprising component $i$ were modeled as bead--spring chains of $M_i$ spherical beads (monomers), each having diameter
$\sigma$ and mass $m$. The interactions between the monomers were purely repulsive and modeled using the
Weeks--Chandler--Andersen potential,\cite{weeks:jcp:1971}
\begin{align}
    U_{\rm m}(r) =
    \begin{cases}
    4 \varepsilon \left[\left(\dfrac{\sigma}{r}\right)^{12} - \left(\dfrac{\sigma}{r}\right)^6\right]
    + \varepsilon, & r \leq 2^{1/6}\sigma
    \\
    0 , & r > 2^{1/6}\sigma
    \end{cases},
    \label{eq:UWCA}
\end{align}
where $r$ is the center-to-center distance between a pair of monomers and $\varepsilon$ sets
the energy scale for the repulsion. Bonded monomers additionally interacted through a
finitely extensible nonlinear elastic potential,\cite{bishop:jcp:1979}
\begin{align}
    U_{\rm b}(r) =
    \begin{cases}
    -\dfrac{k r_0^2}{2} \ln\left[1-\dfrac{r^2}{r_0^2}\right] , & r \leq r_0  \\
    \infty , & r > r_0
    \end{cases}
    \label{eq:UFENE},
\end{align}
with spring constant $k=30\,\varepsilon/\sigma^2$ and maximum bond length $r_0=1.5\,\sigma$ to
prevent unphysical chain crossing.\cite{kremer:jcp:1990} These interactions correspond to good
solvent conditions for dilute polymer solutions.

A droplet with radius $R$ was created by confining the monomers to a spherical domain using the
repulsive part of a harmonic potential,\cite{pieranski:prl:1980}
\begin{align}
    U_{\rm d}(\mathbf{r}) =
    \begin{cases}
    0 , & |\mathbf{r}| \le R-\sigma/2 ,
    \\
    \varepsilon_{\rm d} \left(|\mathbf{r}|-R+\sigma/2\right)^2, & |\mathbf{r}| > R-\sigma/2
    \end{cases},
    \label{eq:UWall}
\end{align}
where $|\mathbf{r}|$ is the distance of a monomer at position $\mathbf{r}$ from the center of the
sphere, and $\varepsilon_{\rm d}$ controls the strength of the repulsion (indirectly, the surface
tension). We chose $\varepsilon_{\rm d} = 100\,\varepsilon/\sigma^2$ to ensure all monomers stayed
fully immersed within the droplet. Equation~\eqref{eq:UWall} assumes a contact angle of $0^\circ$
for the monomers with the interface and neglects any capillary attractions between monomers at the
interface. It also enforces a spherical shape at all times, which may impede deformation or buckling
during the late stages of drying.

Evaporation was mimicked by reducing the droplet radius as a function of time $t$. For a free liquid
droplet, the rate of drying is limited by diffusion of the solvent through the surrounding air, so
the time-dependent droplet radius is,\cite{langmuir:pr:1918, liu:acsnano:2019}
\begin{align}
	R^2= R_0^2 - \frac{\alpha}{4\pi} t,
\end{align}
where $R_0$ is the initial droplet radius and $\alpha$ is the rate of change of surface area that
depends on physical properties such as the solvent's vapor--liquid coexistence densities and
vapor diffusivity. The corresponding receding speed of the interface $v$ is
\begin{align}
	v = -\frac{{\rm d}R}{{\rm d}t} = \frac{\alpha}{8\pi R} .
\end{align}
This drying model neglects potential decreases in the rate of evaporation due to the formation of a
skin layer at the interface \cite{Okuzono:2006gf}. However, this approximation should have only a minor
impact on our results, as recent experiments of drying binary colloidal droplets found that $\alpha$
was nearly constant until the droplet was completely dried.\cite{liu:acsnano:2019}

HI between polymers in the droplet were treated approximately using the
multiparticle collision dynamics (MPCD) technique.\cite{malevanets:jcp:1999, gompper:adv:2009, howard:coce:2019}
The solvent was modeled explicitly as point particles with mass $m_0$ whose motion was governed
by alternating streaming and collision steps. In the streaming step, the
solvent particles were moved ballistically for a period of time $\Delta t_0$. Then, the
solvent particles and monomers were sorted into cubic cells with edge length $\sigma$, subject
to a random shift to ensure Galilean invariance of the algorithm.\cite{ihle:pre:2001} A
momentum-exchanging collision step was then performed between particles in the same cell using the
stochastic rotation dynamics (SRD) variant of MPCD, where the velocity of each particle
relative to its cell's center-of-mass velocity is rotated by a fixed angle around a randomly oriented axis \cite{malevanets:jcp:1999}. This procedure
locally conserves momentum and energy, approximately reproducing HI down to the size of a collision
cell.\cite{huang:pre:2012, dahirel:pre:2018} Because thermodynamically consistent
MPCD algorithms for multiphase systems have only
recently been proposed \cite{Eisenstecken:2018dd}, we neglected any interactions of the solvent
particles with the droplet interface, i.e., the solvent freely flowed through the interface. Hence,
the HI were treated as if the polymers were confined to a spherical domain within a bulk fluid,
e.g., by a semipermeable membrane. We expect this approximate treatment of the solvent boundary
conditions to overpredict solute entrainment \cite{AponteRivera:2016bm} but to be a significant improvement
over complete neglect of HI \cite{Statt:2018bw}.

A droplet having initial radius $R_0 = 50\,\sigma$ was filled with $N_{\rm S} = 2440$
short polymers ($M_{\rm S} = 10$) and $N_{\rm L} = 305$ long polymers ($M_{\rm L} = 80$).
Hence, there were $N_{\rm m} = N_{\rm S}M_{\rm S} + N_{\rm L}M_{\rm L} = 48\,800$ monomers at an
initial monomer volume fraction of $\phi_{\rm m} = N_{\rm m} \sigma^3/(8R_0^3) \approx 0.05$.
Further, the fraction of monomers belonging to long polymers was $\chi = N_{\rm L}M_{\rm L}/N_{\rm m}
= 0.5$. The spherical droplet was placed at the center of a cubic simulation box with edge length
$200\,\sigma$, and the entire box was filled with solvent particles at number density $\rho_{\rm 0}
= 5\,\sigma^{-3}$ ($N_0 = 4 \times 10^7$ solvent particles). Periodic boundary conditions were
employed for the solvent in all Cartesian directions, which can introduce finite-size artifacts due
to coupling of long-ranged HI between periodic images \cite{Yeh:2004gs}; we chose the box edge length to be twice the droplet diameter,
which was the largest size that was computationally feasible for us, to mitigate these
effects. The mass of the monomers was set to the average mass of a cell filled with only solvent
($m = 5\,m_0$), the SRD collision angle was $130^\circ$, and the temperature of the
solution was held constant at $T = 1.0\,\varepsilon/k_{\rm B}$ using a cell-level Maxwellian
thermostat.\cite{huang:jcp:2010} The solvent collision time was $\Delta t_0 = 0.1\,\tau$,
with $\tau = \sqrt{m_0\sigma^2/\varepsilon}$ being the unit of time. The motion of the polymers
between stochastic collisions was integrated using a Verlet scheme with time step $\Delta t =
0.005\,\tau$. All simulations were performed on graphics processing units using HOOMD-blue
(version 2.8.1) \cite{Anderson:2008vg,Glaser:2015cu,howard:cpc:2018} with functionality extended using azplugins (version 0.9.0) \cite{azplugins}.
Unless stated otherwise, all simulations were performed with double-precision floating-point arithmetic
to improve numerical accuracy.

To clearly elucidate the role of HI, we performed implicit-solvent
Langevin dynamics (LD) simulations in addition to the MPCD simulations. This approach allows us to use the {\it same}
polymer model so that any differences between the MPCD and LD simulation results can
be traced back to the treatment of the solvent-mediated hydrodynamics. LD simulations model free-draining HI
and neglect hydrodynamic coupling between monomers. The polymer self-diffusion coefficient $D_i$ in dilute solution
then follows Rouse scaling \cite{Doi,Teraoka:2002up},
\begin{equation}
	D_i = \frac{k_{\rm B}T}{\xi_i M_i},
	\label{eq:DRouse}
\end{equation}
with monomer friction coefficient $\xi_i$. In contrast, the MPCD simulations include hydrodynamic
coupling, and $D_i$ instead follows Zimm scaling \cite{Doi,Teraoka:2002up,mussawisade:jcp:2005},
\begin{equation}
	D_i \approx \frac{k_{\rm B}T}{\xi_i M_i^\nu},
	\label{eq:DZimm}
\end{equation}
where $\nu$ is the fractal dimension of the polymer ($\nu \approx 0.588$ for linear chains in a good solvent).

To approximately match the long-time diffusive dynamics of the polymers between the MPCD and
LD simulations, we followed our previous approach \cite{Howard:2018fb,Statt:2018bw} and adjusted $\xi_i$ in the LD simulations so that
we obtained the same value of $D_i$ in dilute solution as in the MPCD simulations.
We first performed MPCD simulations of pure polymer solutions ($\chi = 0.0$ or $1.0$)
in cubic simulation boxes with edge length $80\,\sigma$. The monomer volume fraction was set to $\phi_{\rm m} = 0.01$, leading to monomer
concentrations $\rho_{\rm m}$ well below the overlap concentration $\rho_{\rm m}^*$ for both
the short and long polymers ($\rho_{\rm m}/\rho_{\rm m}^* \approx 0.04$ and $0.26$, respectively).
We computed $D_i$ from the mean squared displacement of the polymer centers of mass during simulations
of length $10^5\,\tau$. We performed three independent simulations for the short polymers and five independent
simulations for the long polymers to improve statistics, finding
$D_{\rm S} = 1.08 \times 10^{-2}\,\sigma^2/\tau$ and $D_{\rm L} = 2.7 \times 10^{-3}\,\sigma^2/\tau$
with measurement uncertainties of $0.2\%$ and $5\%$, respectively, based on the standard error of the mean.
This measurement, combined with Eq.~\eqref{eq:DRouse}, gave $\xi_{\rm S} = 9.3\,m_0/\tau$ and
$\xi_{\rm L} = 4.6\,m_0/\tau$ for the monomer friction coefficients in the LD simulations.
In what follows, we will indicate results from the MPCD simulations (including HI) as ``+HI'',
whereas results from the LD simulations (without HI) will be indicated as ``--HI''.

\section{Results}
\label{sec:results}

\subsection{Equilibrium properties}
\label{sec:equilibrium}
We first confirmed that the +HI and --HI simulations gave similar equilibrium properties for bulk
polymer solutions. We conducted +HI and --HI simulations at various monomer volume fractions
($0.05 \leq \phi_{\rm m} \leq 0.20$) and compositions ($0.0 \leq \chi \leq 1.0$), using
a similar protocol as for fitting $\xi_i$ except that we now performed only one
simulation per state point. We characterized the conformation of the polymers through the
average gyration tensor,
\begin{align}
	\vv{G} = \frac{1}{M} \sum_{i=1}^M \langle \Delta \vv{r}_i \Delta \vv{r}_i^T \rangle
	\label{eq:G}
\end{align}
where $\Delta \vv{r}_i$ is the vector to monomer $i$ from the polymer center of mass.
The radius of gyration was taken as $R_{\rm g} = (G_{xx} + G_{yy} + G_{zz})^{1/2}$.
As expected, we found that the polymers were isotropic and that the +HI and --HI simulations
produced the same $R_{\rm g}$. We determined $R_{\rm g,L} \approx 6.1\,\sigma$
for the long polymers at $\phi_{\rm m} = 0.05$, which decreased to $R_{\rm g,L} \approx 5.4\,\sigma$
at $\phi_{\rm m} = 0.20$. The short chains shrank less than the long chains, having
$R_{\rm g,S} \approx 1.7\,\sigma$ at $\phi_{\rm m} = 0.05$ and $R_{\rm g,S} \approx 1.6\,\sigma$
at $\phi_{\rm m} = 0.20$. In all cases, the size of the chains was
essentially independent of the composition $\chi$ at constant $\phi_{\rm m}$.

\begin{figure}[!h]
	\includegraphics{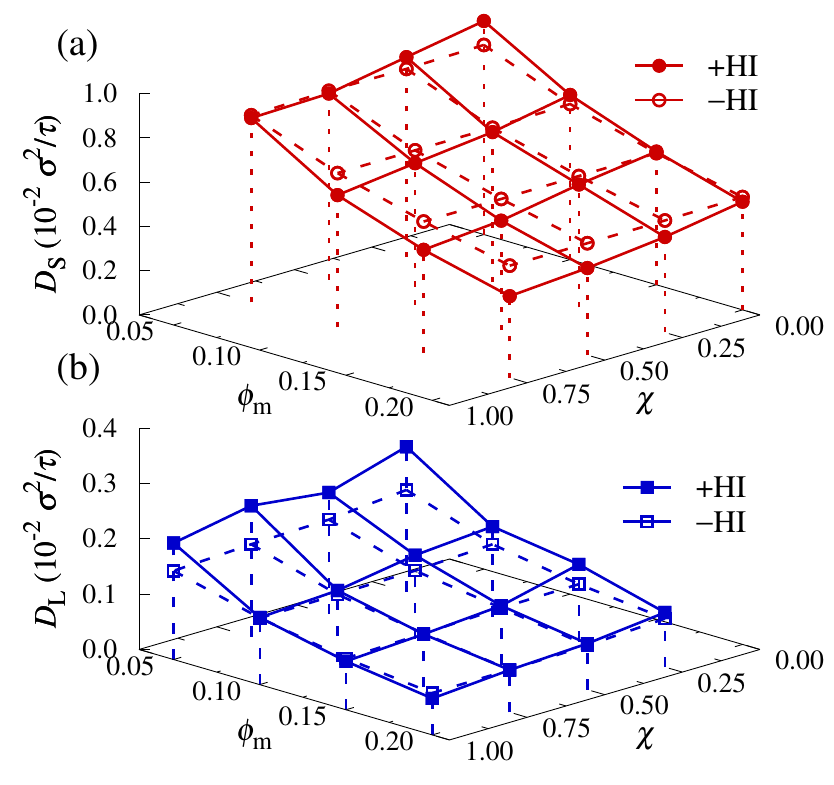}
	\caption{Self-diffusion coefficients for the (a) short polymers $D_{\rm S}$ (red circles) and
	(b) long polymers $D_{\rm L}$ (blue squares) in bulk solutions as a function of monomer volume
	fraction $\phi_{\rm m}$ and composition $\chi$ with HI (filled symbols) and without HI (open symbols).
	\label{fig:bulkDiff}
	}
\end{figure}

Figure~\ref{fig:bulkDiff} compares the bulk self-diffusion coefficients of the short ($D_{\rm S}$) and
long ($D_{\rm L}$) polymers as a function of $\phi_{\rm m}$ and $\chi$ for the +HI and --HI
simulations. In both cases, $D_{\rm S}$ and $D_{\rm L}$ decreased with increasing $\phi_{\rm m}$, as
expected from the concomitant increase in solution viscosity. Interestingly, changing the
composition $\chi$ had a much weaker effect at fixed $\phi_{\rm m}$. For the
short chains, $D_{\rm S}$ agreed well between the +HI and --HI simulations for the short-polymer solutions
($\chi=0.0$), and the agreement generally improved with increasing $\phi_{\rm m}$, where HI are increasingly
screened. Deviations between the +HI and --HI results became more pronounced as the fraction of long
chains $\chi$ increased, but the overall agreement was good (maximum deviation of $25\%$) for all
simulated compositions. For the long chains, the +HI and --HI simulations matched well at high
$\phi_{\rm m}$, and the agreement was better when the mixture had fewer short chains (larger $\chi$). Despite
differences at some compositions, the overall agreement between the +HI and --HI data was again reasonable
(maximum deviation of $35\%$).

\subsection{Evaporation}
\label{sec:evap}
Having established the equilibrium properties of the polymer solutions and good consistency between the
+HI and --HI models, we then performed nonequilibrium simulations of the solutions in drying droplets.
The evaporation process is typically characterized by the droplet P{\'e}clet numbers, ${\rm Pe}_i$, which
describe the relative contributions of advection and diffusion to the motion of the
polymers \cite{Routh:2004jz,Trueman:2012er}. For the spherical geometry employed in this work, we defined ${\rm Pe}_i = v_0 R_0/D_i$
using the initial droplet radius $R_0$ as the characteristic length scale and the
initial evaporation speed $v_0 = \alpha/(8\pi R_0)$ as the characteristic velocity.
Note that ${\rm Pe}_{\rm S}$ and ${\rm Pe}_{\rm L}$ differ for the short and long polymers due to their
different diffusion coefficients and ${\rm Pe}_{\rm L} > {\rm Pe}_{\rm S}$.
We conducted simulations at six drying speeds (Table \ref{tab:params}) and repeated
each simulation eight times for both the +HI and --HI models to improve statistics. For
computational efficiency, the +HI drying simulations used mixed-precision arithmetic \cite{howard:cpc:2018}
and the --HI simulations used single-precision arithmetic.

\begin{table}[!h]
	\centering
	\caption{Initial evaporation speed $v_0$ (with corresponding rate of change of surface area $\alpha$)
	and P{\'e}clet numbers for the short polymers ${\rm Pe}_{\rm S}$ and long polymers ${\rm Pe}_{\rm L}$
	in a droplet with initial radius $R_0 = 50\,\sigma$.}
	\label{tab:params}
	\begin{tabular}{cccc}
	\hline
	$v_0$ ($\sigma/\tau$) & $\alpha$ ($\sigma^2/\tau$) & ${\rm Pe}_{\rm S}$ & ${\rm Pe}_{\rm L}$ \\
	\hline
	0.0001 & 0.13 & 0.46 & 1.8 \\
	0.0005 & 0.63 & 2.3 & 9.2 \\
	0.001 & 1.3 & 4.6 & 18 \\
	0.002 & 2.5 & 9.3 & 37 \\
	0.005 & 6.3 & 23 & 92 \\
	0.01 & 13 & 46 & 180 \\
	\hline
	\end{tabular}
\end{table}

\begin{figure*}[!tb]
	\includegraphics{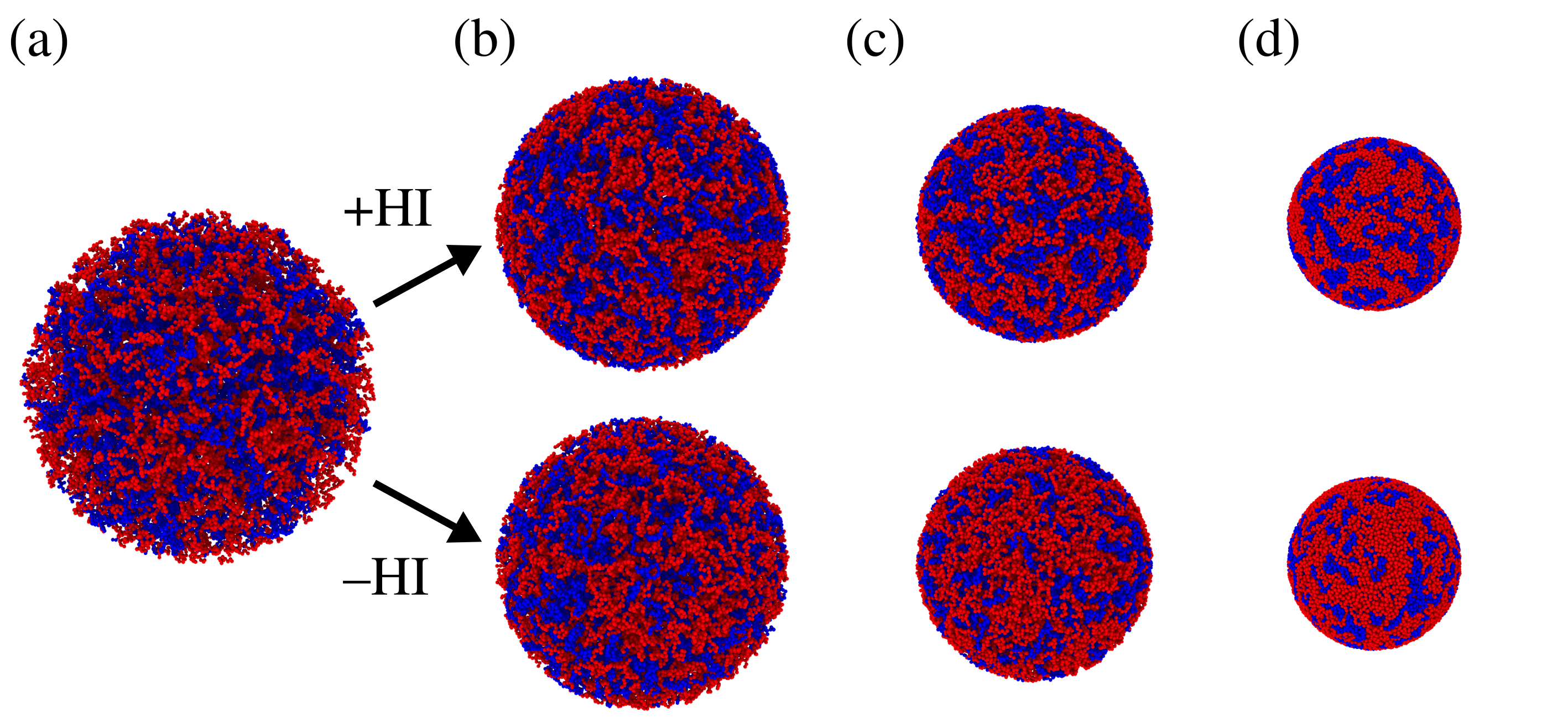}
	\caption{Simulation snapshots showing the outside of the drying droplet at
	(a) $R = 50.0\,\sigma$, (b) $41.8\,\sigma$, (c) $33.9\,\sigma$, and (d) $25.0\,\sigma$
	in the +HI (top) and --HI (bottom) simulations. The initial evaporation speed was
	$v_0 = 0.002\,\sigma/\tau$. The short and long polymers are colored red and blue, respectively.}
	\label{fig:snapshots}
\end{figure*}

\begin{figure}[!h]
	\includegraphics{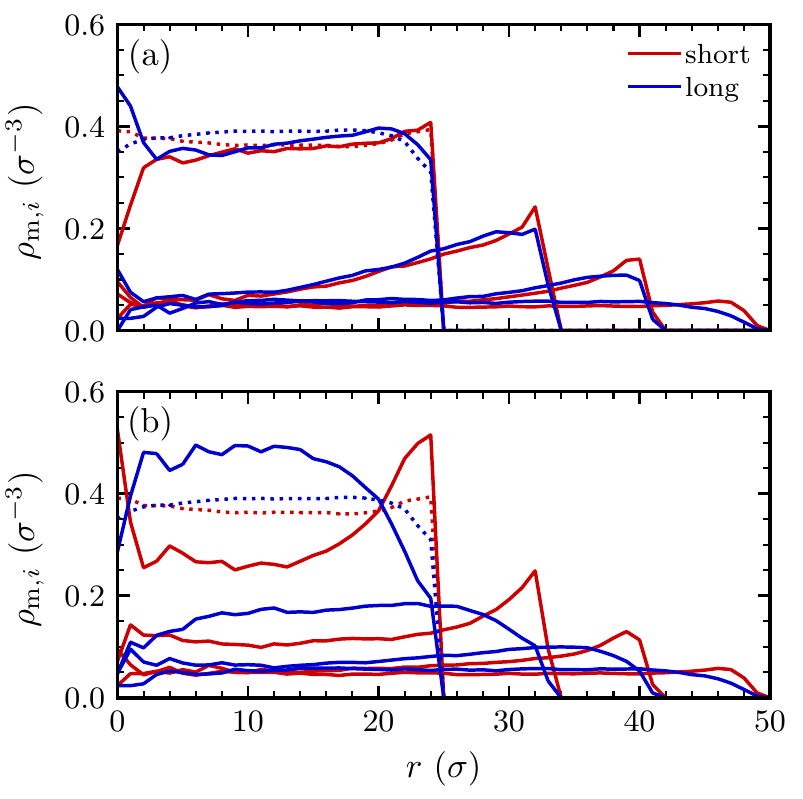}
	\caption{Radial monomer density profiles $\rho_{{\rm m},i}$ of the short polymers (red) and long
	polymers (blue) from the (a) +HI and (b) --HI simulations for the same $v_0$ and $R$
	as in Fig.~\ref{fig:snapshots}. The equilibrium profiles for $R = 25.0\,\sigma$ are also shown as dotted lines.}
	\label{fig:density.v0.002}
\end{figure}

Figure \ref{fig:snapshots} shows snapshots of the drying droplet from one of the +HI and --HI
simulations for $v_0 = 0.002\,\sigma/\tau$. These snapshots reveal that the mixture formed
a weakly stratified core--shell structure in the --HI simulations,
with the shorter polymers enhanced near the droplet--air interface. It is, however, difficult to
visually judge the extent of stratification from these snapshots due to the interpenetration
of the polymers. To better assess the presence of stratification, we computed the corresponding
average radial density profiles of monomers from the short and long polymers $\rho_{{\rm m},i}$.
Figure \ref{fig:density.v0.002} shows these profiles for the same drying rate and time points as
in Fig.~\ref{fig:snapshots}, averaged over the eight simulations. The profiles confirm the formation of a core--shell morphology
in the --HI simulations, but the morphology in the +HI simulations remained essentially homogeneous.
To verify that the stratified structures in the --HI simulations were a result of the drying, we
continued the simulations with the final droplet radius held constant at $R=25\,\sigma$, and indeed
observed that the monomer distributions relaxed to (almost) uniform distributions (dotted lines
in Fig.~\ref{fig:density.v0.002}).

\begin{figure}[!h]
	\includegraphics{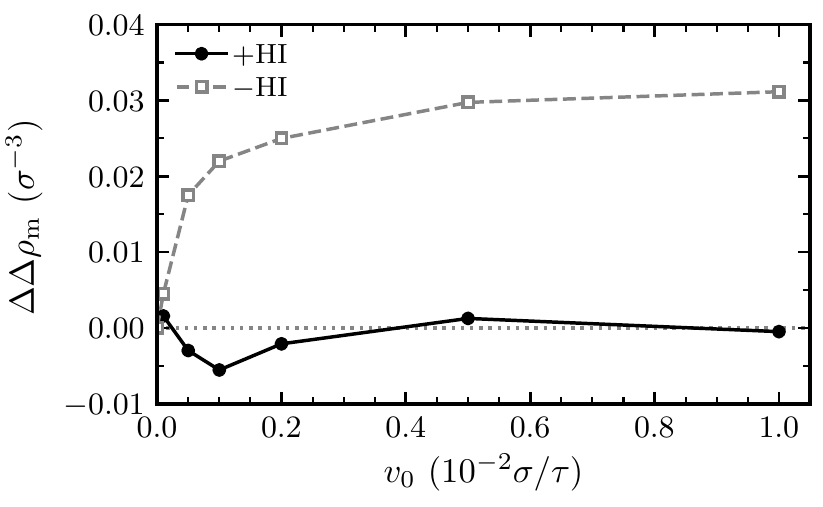}
	\caption{Order parameter $\Delta \Delta \rho_{\rm m}$ as a function of
	initial evaporation speed $v_0$ for the +HI and --HI simulations.
	$\Delta\Delta \rho_{\rm m} > 0$ indicates small-on-top stratification,
	while $\Delta\Delta \rho_{\rm m} < 0$ indicates more long polymers near the
	surface compared to the equilibrium distribution.}
	\label{fig:orderParameter}
\end{figure}

In order to quantify the degree of stratification near the droplet surface,
we computed the average difference in the densities of monomers from the
short and long chains in a thin shell comparable to the size of the large polymer,
\begin{align}
	\Delta \rho_{\rm m} = \frac{3}{R^3-R'^3} \int_{R'}^R \left[ \rho_{\rm m, S}(r)-\rho_{\rm m, L}(r) \right] r^2{\rm d}r,
	\label{eq:deltaN}
\end{align}
where $R' = R-2 R_{\rm g,L}$ and $R_{\rm g,L} \approx 6.1\,\sigma$
is the radius of gyration of the long chains at the initial monomer volume fraction
$\phi_{\rm m} = 0.05$. The case $\Delta \rho_{\rm m} > 0$ indicates an excess of monomers from short
polymers, whereas $\Delta \rho_{\rm m} < 0$ indicates the opposite. In a perfectly homogeneous
solution at composition $\chi=0.5$, we expect $\Delta \rho_{\rm m} = 0$. However, even in equilibrium,
there is a small (but noticeable) excess of short chains close to the droplet--air interface due
to entropic effects (Fig. \ref{fig:density.v0.002}). To remove this inherent offset, we
computed $\Delta \rho_{\rm m, eq}$ at the final droplet radius ($R=25\,\sigma$) after the
mixture equilibrated and subtracted it from $\Delta \rho_{\rm m}$. Figure~\ref{fig:orderParameter} shows the resulting order
parameter $\Delta\Delta \rho_{\rm m} = \Delta \rho_{\rm m} - \Delta \rho_{\rm m, eq}$ as a function of
initial evaporation speed $v_0$ for both the +HI and --HI simulations. These data clearly show that the polymer
mixtures remained essentially homogeneous in the +HI simulations for all investigated evaporation
speeds. In contrast, there was distinct stratification in the --HI simulations, which
became more pronounced as the droplets dried faster.

\begin{figure}[!h]
	\includegraphics{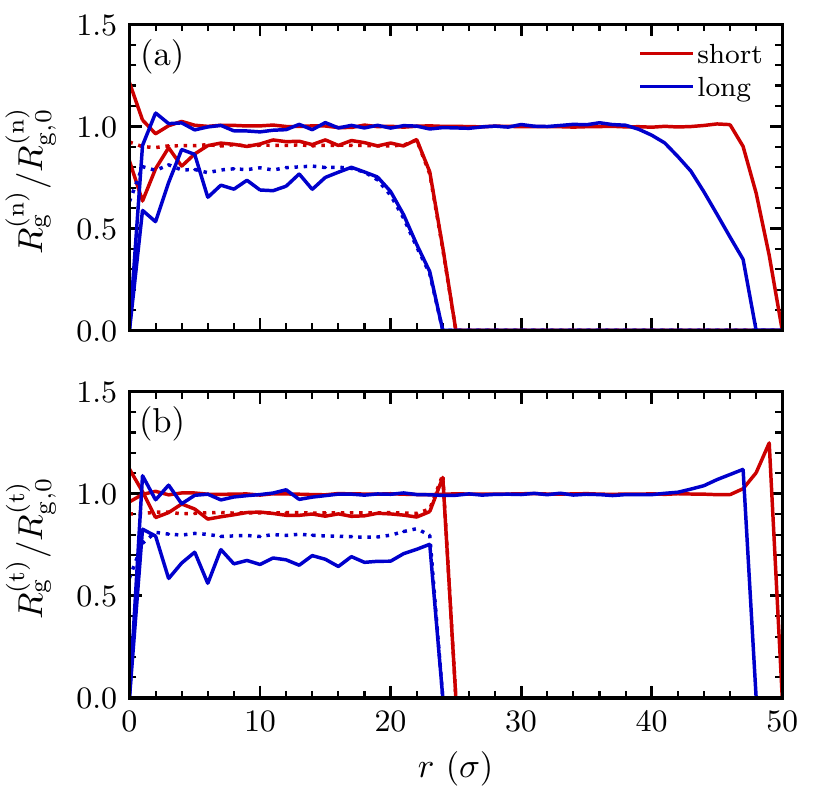}
    \caption{Radial profiles of the (a) normal component $R_{\rm g}^{({\rm n})}$ and
    (b) tangential component $R_{\rm g}^{({\rm t})}$ of the radius of gyration
    for the short polymers (red) and long polymers (blue) before ($R = 50.0\,\sigma$)
    and after ($R=25.0\,\sigma$) drying at $v_0 = 0.002\,\sigma/\tau$ in the +HI
    simulations. The values are normalized by their bulk values at the initial volume fraction
    and composition ($\phi_{\rm m} = 0.05$ and $\chi = 0.5$), i.e.,
    $R_{{\rm g},0}^{({\rm n})} = R_{{\rm g},0}/\sqrt{3}$ and
    $R_{{\rm g},0}^{({\rm t})} = R_{{\rm g},0}\sqrt{2/3}$ with $R_{{\rm g},0}$
    being the bulk value of the radius of gyration discussed in Section~\ref{sec:equilibrium}.
    The equilibrium profiles for $R = 25.0\,\sigma$ are also shown as dotted lines.
    Profiles after drying in the --HI simulations were essentially the same and have
    been omitted for clarity.}
	\label{fig:rg}
\end{figure}

Despite these differences in overall morphology, we found that the conformations of
the polymers within the droplet were essentially the same in both the +HI and --HI
simulations. We computed the gyration tensor of the polymers in the droplet before drying and
at the end of drying using Eq.~\eqref{eq:G} and determined radial profiles by
averaging according to the position of the polymer centers of mass. We then
computed the normal component $R_{\rm g}^{({\rm n})}$ and tangential component
$R_{\rm g}^{({\rm t})}$ of the total radius of gyration $R_{\rm g}$ relative to
the droplet surface. Figure~\ref{fig:rg} shows these components for the +HI simulations
normalized relative to their values in bulk at the initial volume fraction and
composition ($\phi_{\rm m} = 0.05$ and $\chi = 0.5$). Before drying, both the long
and short polymers had isotropic shapes ($R_{\rm g}^{({\rm n})} = R_{\rm g}^{({\rm t})}/\sqrt{2}$)
except near the surface of the droplet, where
both tended to be slightly stretched tangential to the surface and compressed normal to the
surface. The long polymers were affected over a longer distance from the surface,
as expected; these effects are fully consistent with equilibrium depletion of the
long polymer near the surface (Fig.~\ref{fig:density.v0.002}). Both
polymers shrank as the mixture concentrated during drying, with the increased
density affecting the long polymers more than the short polymers (as in
the equilibrium bulk solutions, see Section~\ref{sec:equilibrium}). However, both
components of $R_{\rm g}$ for the short polymers stayed close to their equilibrium
profiles (i.e., those obtained when we stopped the drying simulations at $R = 25.0\,\sigma$ and relaxed
the mixture, dotted line), while the long polymers tended to be slightly smaller
after drying compared to equilibrium.
Although Fig.~\ref{fig:rg} shows profiles for the +HI simulations, the
--HI simulations had nearly identical final profiles for the components of $R_{\rm g}$.
This may be partially due to the fact that the total monomer densities are similar
in both simulations even though the local polymer compositions are drastically different
(Fig.~\ref{fig:density.v0.002}). The most striking differences in structure were
obtained in the (mesoscopic) droplet morphology rather than the (microscopic)
polymer conformations, with HI between polymers seemingly playing an important role
in setting the droplet morphology.

\subsection{Onsager coefficients}
\label{sec:onsager}
The qualitatively different morphologies obtained in the +HI and --HI drying
simulations are completely consistent with prior simulations that compared
stratification of polymer solutions in drying films with and without
HI \cite{Statt:2018bw}. It is then clearly important to include HI, which are
inherent to experiments, to build predictive models for this dynamic process.
However, this task can be computationally demanding even with a mesoscale simulation
technique like MPCD. Continuum multicomponent diffusion models (Eq.~\eqref{eq:onsager})
remain attractive in this respect, but the solvent must be properly accounted
for in the thermodynamic model through the chemical potentials $\mu_i$ and in
the transport model through the phenomenological Onsager coefficients $L_{ij}$.
Unlike prior work \cite{Statt:2018bw,Howard:2018fb,Tang:2019jy}, our polymer model has
identical thermodynamics in both the +HI and --HI simulations, so the polymer chemical
potentials $\mu_{\rm S}$ and $\mu_{\rm L}$ are the same in both. Comparison of
the two simulations then allows us to isolate the role of the solvent in the process and how
HI impact $L_{ij}$.

To facilitate this comparison, it will be preferable to convert the Onsager coefficients
$L_{ij}$ defined by Eq.~\eqref{eq:onsager} to effective phenomenological coefficients
$\Lambda_{ij}$ defined by Eq.~\eqref{eq:effonsager}. This conversion is necessary because the
+HI simulations model a true three-component mixture, whereas the --HI simulations model an
effective two-component mixture. In the --HI simulations, the implicit solvent
is a stationary background whose properties do not depend on the polymers,
so it is trivially the case that $\Lambda_{ij} = L_{ij}$.
In the +HI simulations, the motion of the solvent is coupled to the motion of
the polymers because linear momentum is conserved locally so that a net force on
(and flux of) the polymers implies a counterforce on (and backflow of) the solvent
in the opposite direction. As a result, the mass-averaged velocity,
\begin{equation}
\bar{\vv{u}}_{\rm m} = \frac{\sum_{i=0}^n m_i \rho_i \vv{u}_i}{\sum_{i=0}^n m_i \rho_i}
\label{eq:ucom}
\end{equation}
(with $m_i$ being the mass of component $i$),
is zero if the net force on the mixture is zero, making $\bar{\vv{u}}_{\rm m}$
an appropriate reference velocity for defining $\vv{j}_i$ in the +HI simulations.
We will assume that the Gibbs--Duhem relationship between the chemical potentials at
constant temperature and pressure is satisfied in the +HI simulations as a
result of the local momentum conservation and zero net force, and also that the
Onsager reciprocal relations hold. With these assumptions and using $\bar{\vv{u}}_{\rm m}$
in Eq.~\eqref{eq:onsager}, the effective phenomenological coefficients are obtained as
\begin{equation}
\Lambda_{ij} = \sum_{k=1}^n L_{ik} \left(\delta_{kj} + \frac{m_k \rho_j}{m_0 \rho_0}\right)
\label{eq:lamL}
\end{equation}
with $\delta_{kj}$ being the Kronecker delta. This relationship fully eliminates the
solvent from the phenomenological model. With Eq.~\eqref{eq:effonsager},
the coefficients $\Lambda_{ij}$ allow for direct comparison of how the polymer
fluxes depend on gradients of the polymer chemical potentials between the
+HI and --HI simulations.

We aimed to use simulations to clarify how HI contribute to $\Lambda_{ij}$
and to test assumptions made in prior models. In particular, the coefficient
$\Lambda_{\rm LS}$ that couples the flux of the long polymers to the chemical
potential gradient of the short polymers contributes to diffusiophoretic
models for stratification like Eq.~\eqref{eq:dp} \cite{Sear:2017cv,Sear:2018dn}, but it was neglected in models
derived from dynamic DFT \cite{Howard:2017bq,Howard:2017vu,Zhou:2017gg}, where $\Lambda_{\rm LL}$ was assumed to be dominant. It is unclear which coefficients are required to properly capture the dynamics
in concentrated solutions of deformable polymers, for which it is not straightforward
to compute the diffusiophoretic coefficient \cite{RamirezHinestrosa:2020gd}.

To this end, we measured $\Lambda_{ij}$ directly in bulk polymer mixtures using
nonequilibrium simulations \cite{Maginn:1993fq}. The initial configurations were taken
from the equilibrium simulations (Sec.~\ref{sec:equilibrium}). We applied a constant
force $\vv{F}_j = F_j \hat{\vv{x}}$ to either the short polymers or the long polymers,
where $\hat{\vv{x}}$ is a unit vector pointing along the $x$ direction. (In practice, this
was achieved by distributing an equal force $(F_j/M_j)\hat{\vv{x}}$ to each bead in the
chain.) The constant force plays the role of a linear chemical potential gradient,
$\nabla \mu_j = -\vv{F}_j$, without requiring a concentration gradient. In
the +HI simulations, a counterforce $\vv{F}_0 = -(N_j/N_0) \vv{F}_j$
was applied to the solvent particles to ensure that the entire
system was force-free, so there were effectively both polymer and solvent
chemical potential gradients in opposite directions.
Simulations were run at multiple values of
$F_{\rm S} \le 0.1\,\varepsilon/\sigma$ and $F_{\rm L} \le 0.4\,\varepsilon/\sigma$
that were sufficiently small that the
polymers did not deform significantly ($G_{xx}$ extended less than $3\%$
relative to its equilibrium value). The composition was fixed at $\chi=0.5$ and the monomer
volume fraction was varied in the range $0.05 \le \phi_{\rm m} \le 0.20$.
The coefficients $\Lambda_{ij}$ were determined by fitting
the average velocity $\vv{u}_i$ of component $i$ in
a stationary reference frame ($\bar{\vv{u}} = \vv{0}$) according to Eq.~\eqref{eq:effonsager},
\begin{equation}
	\rho_i \vv{u}_i = \Lambda_{ij} \vv{F}_j.
	\label{eq:lamfit}
\end{equation}
We measured the average velocities every $2.5\,\tau$ during a simulation of length
$10^5\,\tau$, and we discarded the first $10\%$ of the data to allow the system to
achieve a steady flux. We computed the average velocities using the remainder of
the data and estimated uncertainties from the standard error between 5 subdivided
blocks of the data.

In the +HI simulations, the net force on the polymer mixture is zero and linear momentum
is conserved, so the reference frame for measuring $\vv{u}_i$
can be considered both the laboratory frame and the mass-averaged frame given by
$\bar{\vv{u}}_{\rm m}$ (Eq.~\eqref{eq:ucom}).
We confirmed that $\bar{\vv{u}}_{\rm m} \lesssim 10^{-8}\,\sigma/\tau$
for all investigated cases, as expected. In a pure polymer solution, momentum
conservation implies that a polymer flux in one direction must be opposed by a
solvent backflow. (An analogous backflow occurs in colloidal suspensions, which
are usually assumed to be incompressible, or to have no net volume flux,
so a flux of colloids implies a solvent backflow.)
In a two-polymer mixture with solvent, a flux of one of the polymer species can be
balanced by not only the solvent backflow but also potentially a flux of the
other polymer species. The distribution of these fluxes will depend on the interactions
between polymers and the coefficients $\Lambda_{ij}$. We emphasize that the mixtures
remained homogeneous in both the +HI and --HI simulations, and that there was
no net flow in the +HI simulations, supporting the assumption of constant pressure in our analysis
of the +HI simulations.

In the --HI simulations, there is no explicit solvent, and $\vv{F}_j$ imposes
a net force. The polymers still reached a steady-state velocity
because the implicit solvent was dissipative, but unlike the +HI simulations,
there was no explicit backflow of solvent (i.e., $\vv{u}_0 = \vv{0}$).
The force on the polymers in the --HI simulations implies a balancing force on
the implicit solvent to make the mixture force-free. This effect
cannot be easily included in the inhomogeneous drying simulations \cite{Fortini:2016ip,MartinFabiani:2016fj,Makepeace:2017ht,Howard:2017bq,Fortini:2017jl,Howard:2017vu}, but it can be
approximately treated in the bulk measurements of $\Lambda_{ij}$. Assuming
a force-free polymer--implicit-solvent mixture and total momentum conservation,
as in the +HI simulations, the measured flux of the polymers implies a solvent backflow,
\begin{equation}
	\vv{u}_0 = -\frac{m}{m_0 \rho_0}(M_{\rm S} \rho_{\rm S} \vv{u}_{\rm S} + M_{\rm L} \rho_{\rm L} \vv{u}_{\rm L}).
	\label{eq:uslv}
\end{equation}
The --HI simulations can then be regarded as having been conducted in either the stationary
laboratory frame or the frame that moves with the solvent at velocity $\vv{u}_0$.
Velocities measured in the latter moving frame can be shifted into a stationary one
where the solvent flows and $\bar{\vv{u}}_{\rm m} = \vv{0}$ using
Eq.~\eqref{eq:uslv}. To make complete comparison with the +HI simulations, we will
present results for $\Lambda_{ij}$ from the --HI simulations taking
$\vv{u}_i$ both unshifted (as in the evaporation simulations) and shifted to account
for backflow. Even with shifting, though, the polymer fluxes and implied
solvent backflow in the --HI simulations can still differ from the +HI simulations,
as these may depend on the presence of HI.

\begin{figure}
	\includegraphics{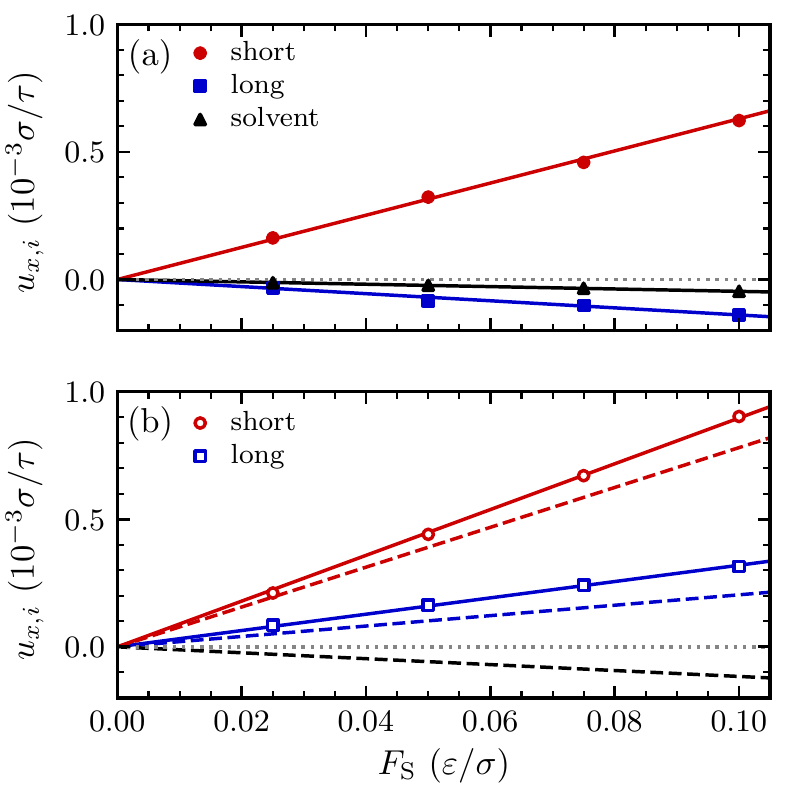}
	\caption{Average velocity $u_{x,i}$ of the short (red circles) and long (blue squares)
	polymers when the short polymers are dragged with a constant force $F_{\rm S}$ in the
	(a) +HI (filled symbols) and (b) --HI (open symbols) simulations at $\phi_{\rm m}=0.10$
	and $\chi=0.5$. In (a), the measured average solvent velocity is also shown
	(black triangles). In (b), the implied solvent velocity was computed using
	Eq.~\eqref{eq:uslv} and is shown as a black dashed line. The polymer velocities
	in this reference frame are also shown as dashed lines. The estimated measurement
	uncertainty is smaller than the symbol size.
	\label{fig:dragPhi0.10}}
\end{figure}

Figure \ref{fig:dragPhi0.10} shows one example of the measured average velocities
$u_{x,i} = \vv{u}_i \cdot \hat{\vv{x}}$ as a function of $F_{\rm S}$ for
$\phi_{\rm m}=0.10$. The relationship between $u_{x,i}$ and
$F_{\rm S}$ was linear, giving another indication that $F_{\rm S}$ was
sufficiently small to reliably perform the measurement. In both the +HI and --HI
simulations, the short polymers migrated in the direction of the applied force, as expected.
In the +HI simulations (Fig.~\ref{fig:dragPhi0.10}a), we measured
not only a solvent backflow ($u_{x,0} < 0$) but also a counterflow
of the long polymers ($u_{x,{\rm L}} < 0$). In contrast, in the --HI simulations
(Fig.~\ref{fig:dragPhi0.10}b), both the short polymers and long polymers migrated
in the same direction ($u_{x,{\rm L}} > 0$) in the laboratory frame. Shifting to account
for the solvent backflow using Eq.~\eqref{eq:uslv} (Fig.~\ref{fig:dragPhi0.10}b, dashed lines) did not
qualitatively alter this behavior but did decrease both $u_{x,{\rm S}}$ and $u_{x,{\rm L}}$.
Analogous behavior was observed in the simulations
having force $F_{\rm L}$ applied to the long polymers with the role of the short and
long polymers exchanged. However, larger $F_{\rm L}$ than $F_{\rm S}$
was required to obtain comparable $u_{x,i}$ due to the decreased mobility of
the long polymers.

\begin{figure*}[!tb]
	\includegraphics{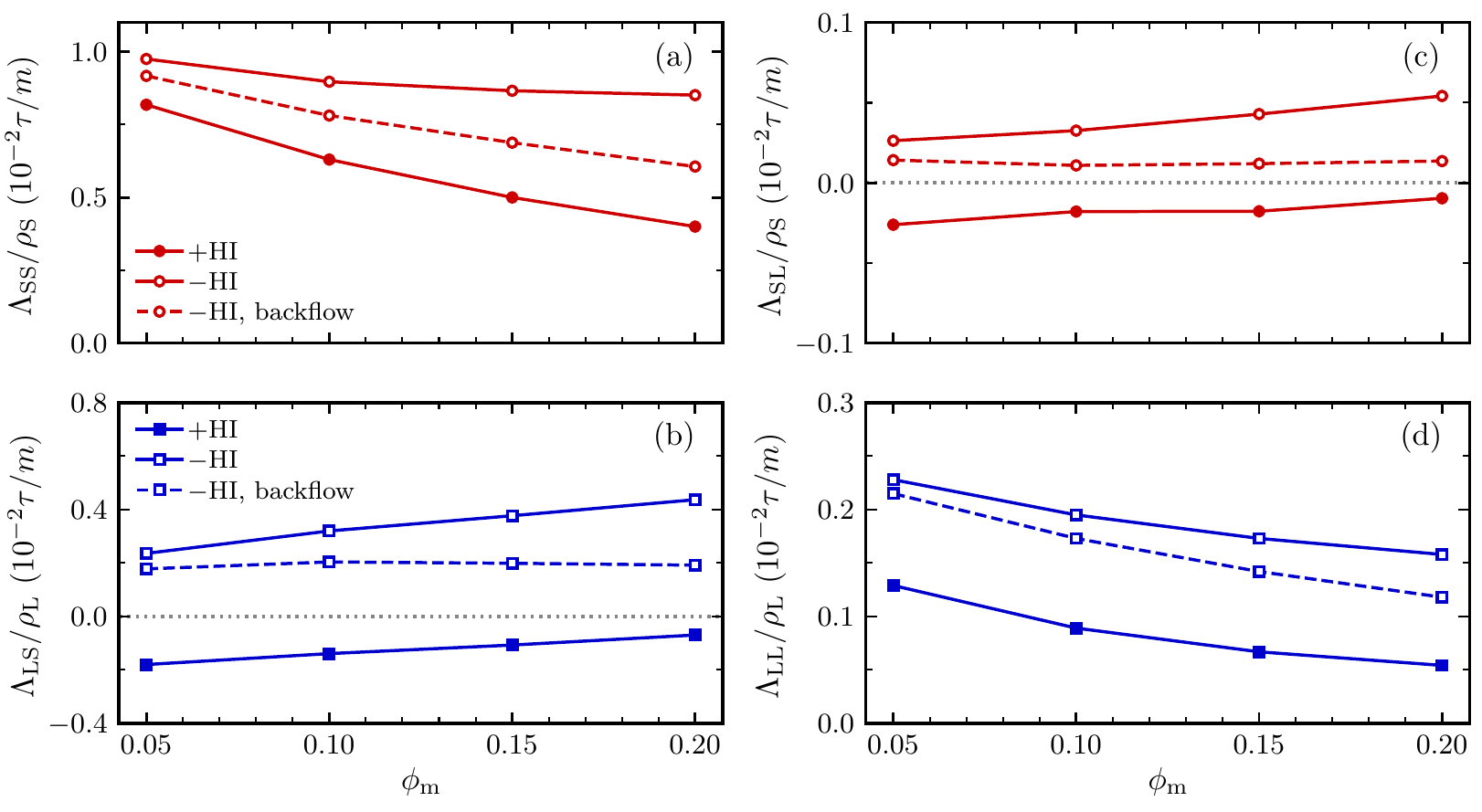}
	\caption{Effective Onsager coefficients (a) $\Lambda_{\rm SS}$, (b) $\Lambda_{\rm LS}$,
	(c) $\Lambda_{\rm SL}$, and (d) $\Lambda_{\rm LL}$ as functions of monomer volume
	fraction $\phi_{\rm m}$ at fixed composition $\chi=0.5$ for the +HI (filled symbols) and
	--HI simulations (open symbols). (Note the different scales for $\Lambda_{ij}$ in each panel.)
	In the --HI simulations, the coefficients are shown for both the reference frame
	where the solvent is stationary (solid lines) and the reference frame where the
	polymer velocities are shifted to account for solvent backflow (dashed lines) using Eq.~\eqref{eq:uslv}.
	As in Fig.~\ref{fig:dragPhi0.10}, the red circles (top row) indicate coefficients
	for the flux of the short polymers, while the blue squares (bottom row) indicate
	coefficients for the flux of the long polymers. The estimated uncertainties from
	fitting the simulation data to Eq.~\eqref{eq:lamfit} are smaller than the symbol size.
	\label{fig:onsagerCoeff}}
\end{figure*}

\begin{figure}
    \includegraphics{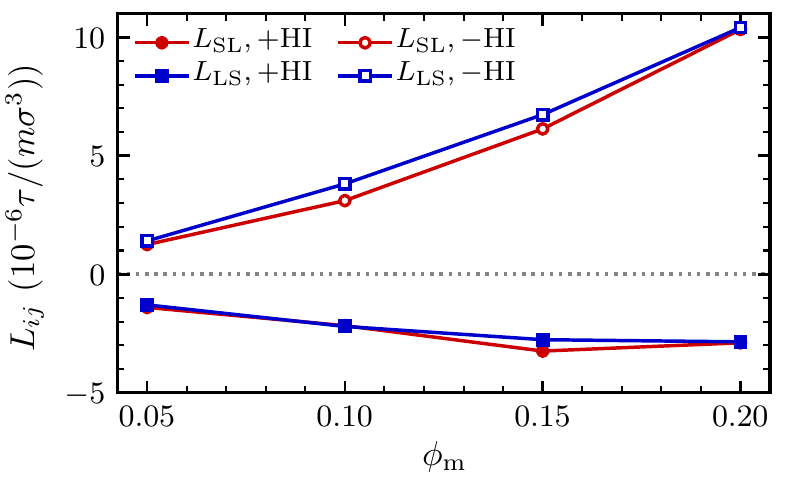}
    \caption{Off-diagonal Onsager coefficients $L_{\rm SL}$ (red circles) and
    $L_{\rm LS}$ (blue squares) as functions of monomer volume fraction
    $\phi_{\rm m}$ at fixed composition $\chi = 0.5$, corresponding to the simulations
    in Fig.~\ref{fig:onsagerCoeff}. The coefficients for the +HI simulations
    (filled symbols) were obtained by inverting Eq.~\eqref{eq:lamL},
    while the coefficients for the --HI simulations (open symbols) are trivially
    $L_{ij} = \Lambda_{ij}$.
    \label{fig:trueonsager}}
\end{figure}

We extracted all four $\Lambda_{ij}$ from simulations with either $F_{\rm S}$ or
$F_{\rm L}$ applied using Eq.~\eqref{eq:lamfit}, shown in Fig.~\ref{fig:onsagerCoeff}.
We first considered the diagonal coefficients $\Lambda_{\rm SS}$ (Fig.~\ref{fig:onsagerCoeff}a)
and $\Lambda_{\rm LL}$ (Fig.~\ref{fig:onsagerCoeff}d) coupling the
flux of the polymers to their own chemical potential gradients. We found that
$\Lambda_{\rm SS}$ and $\Lambda_{\rm LL}$ were both positive and consistently smaller
with HI than without HI, indicating that HI retard the motion of the polymers
as in Batchelor's analysis of sedimenting colloidal suspensions \cite{Batchelor:1976gz}. This effect
was (relatively) more significant for the long polymers (Fig.~\ref{fig:onsagerCoeff}d)
than for the short polymers (Fig.~\ref{fig:onsagerCoeff}a) at low $\phi_{\rm m}$.
The measured $\Lambda_{ii}$ are in modest agreement with the ansatz in the
dynamic DFT models, $\Lambda_{ii}/\rho_i \approx D_i/k_{\rm B}T$, with an error
of roughly 50\% or less for both the +HI and --HI simulations.
However, more significant qualitative differences between the +HI and --HI simulations are apparent
for the off-diagonal coefficients $\Lambda_{\rm LS}$ (Fig.~\ref{fig:onsagerCoeff}b)
and $\Lambda_{\rm SL}$ (Fig.~\ref{fig:onsagerCoeff}c), which have
different signs in the +HI and --HI simulations. The off-diagonal
contributions are positive and \textit{promote} diffusion without HI
but are negative and \textit{inhibit} diffusion with HI. (We also computed the
underlying off-diagonal Onsager coefficients $L_{\rm LS}$ and $L_{\rm SL}$
by inverting Eq.~\eqref{eq:lamL}; they showed the same qualitative differences as $\Lambda_{\rm LS}$
and $\Lambda_{\rm SL}$ and are symmetric, as expected (Fig.~\ref{fig:trueonsager}).)
Inclusion of solvent backflow in the --HI simulations quantitatively shifted all $\Lambda_{ij}$; however,
the same qualitative differences were apparent in the off-diagonal components.
Evidently, not only solvent backflow but also HI between polymers play an
important role in setting $\Lambda_{ij}$ that needs to be considered.

We focus on $\Lambda_{\rm LS}$, which contributes to the diffusiophoretic
coefficient for the long polymers. For all tested $\phi_{\rm m}$,
$\Lambda_{\rm LS} > 0$ in the --HI simulations, but $\Lambda_{\rm LS} < 0$ in the +HI simulations.
If the diagonal contribution $\Lambda_{\rm LL}$ to the long-polymer flux is neglected, the
change in sign of $\Lambda_{\rm LS}$ fully explains the presence or absence of
stratification in the --HI and +HI simulations, respectively. However, the
picture is more complex for nondilute polymer solutions. First, the short-polymer
concentration profile (chemical potential gradient) evolves dynamically and is
determined by both the short-polymer and long-polymer concentrations through
$\Lambda_{\rm SS}$ and $\Lambda_{\rm SL}$ (Fig.~\ref{fig:onsagerCoeff}c). Hence,
small quantitative differences in all $\Lambda_{ij}$ between the +HI and --HI
simulations may contribute to the final morphology. Second, the flux of
the long polymers due to their own chemical potential is nonnegligible because
neither the short polymers nor the long polymers are dilute.
This, in particular, may promote stratification when $\Lambda_{\rm LL} > 0$ even as
$\Lambda_{\rm LS} < 0$ tends to suppress it in the +HI simulations, and
the net effect of these two contributions will depend on the magnitudes of
$\nabla \mu_{\rm S}$ and $\nabla \mu_{\rm L}$.

For the bead--spring polymer model studied, the polymer chemical potentials can be well approximated
using a hard-chain equation of state \cite{Howard:2017vu,Jackson:1988kl,Chapman:1988ge}. We separate the chemical potential
into ideal and excess parts, $\nabla \mu_j = \nabla \mu_j^{\rm id} + \nabla \mu_j^{\rm ex}$.
The ideal term $\nabla \mu_j^{\rm id} = k_{\rm B}T \nabla \ln (\lambda_j^3 \rho_j)$ depends on the
\textit{individual} density of component $j$, while $\nabla \mu_j^{\rm ex}$ is proportional to $M_j$
and depends on the \textit{total} monomer volume fraction $\phi_{\rm m}$ and polymer number density
$\rho_{\rm S}+\rho_{\rm L}$. In a sufficiently dense solution,
the ideal contribution can be neglected and $\nabla \mu_j \approx \nabla \mu_j^{\rm ex}$ to first approximation,
so $\nabla \mu_{\rm L} \approx (M_{\rm L}/M_{\rm S}) \nabla \mu_{\rm S}$. In this
regime, the flux of the short or long polymers can then be approximately computed using a single
effective coefficient, $\vv{j}_i \approx \tilde\Lambda_i \nabla\mu_{\rm S}$, where
\begin{equation}
\tilde\Lambda_i = \Lambda_{i{\rm S}} + \left(\frac{M_{\rm L}}{M_{\rm S}}\right)\Lambda_{i{\rm L}}.
\label{eq:exonsager}
\end{equation}
Small-on-top stratification is usually expected when $|\vv{u}_{\rm L}| > |\vv{u}_{\rm S}|$ so that
the long polymers separate from the short polymers as they codiffuse \cite{Fortini:2016ip,Fortini:2017jl,Howard:2017bq,Howard:2017vu,Zhou:2017gg},
and therefore an approximate condition for stratification in the nondilute regime is
$\tilde\Lambda_{\rm L}/\rho_{\rm L} > \tilde\Lambda_{\rm S}/\rho_{\rm S}$ across
a range of compositions.

\begin{figure}
	\includegraphics{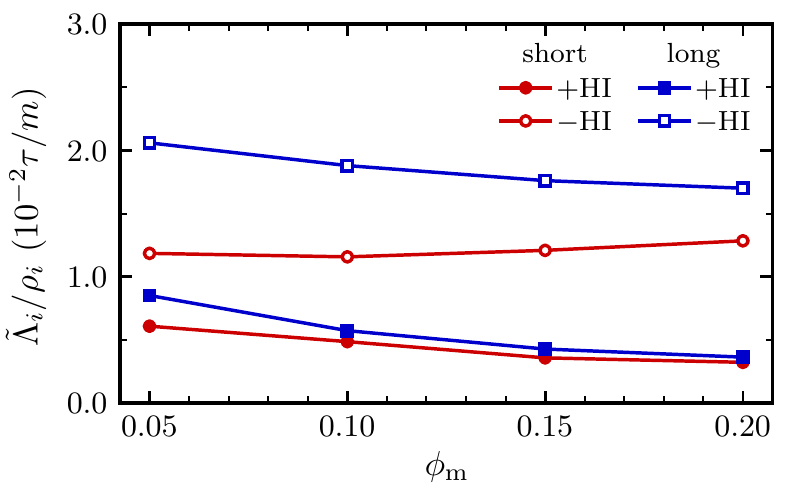}
	\caption{Approximate coefficients $\tilde\Lambda_i/\rho_i$ for the short (red circles)
	and long (blue squares) polymers in the +HI (filled symbols) and --HI (open symbols)
	simulations, computed from Eq.~\eqref{eq:exonsager}.}
	\label{fig:exonsager}
\end{figure}

Figure \ref{fig:exonsager} shows $\tilde\Lambda_i/\rho_i$ for
both the +HI simulations and the --HI simulations in the laboratory frame used
for the evaporation simulations. In the --HI simulations,
$\tilde\Lambda_{\rm L}/\rho_{\rm L} > \tilde\Lambda_{\rm S}/\rho_{\rm S}$ for all
$\phi_{\rm m}$, consistent with the presence of stratification in that model.
In contrast, $\tilde \Lambda_{\rm L}/\rho_{\rm L} \approx \tilde\Lambda_{\rm S}/\rho_{\rm S}$
in the +HI simulations for all $\phi_{\rm m}$, and $\tilde \Lambda_{\rm L}/\rho_{\rm L}$
and $\tilde\Lambda_{\rm S}/\rho_{\rm S}$ are both smaller than in the --HI
simulations, suggesting that the two components will not readily stratify.
Inspection of the relative magnitudes of the diagonal and off-diagonal contributions
to $\tilde\Lambda_i$ (Fig.~\ref{fig:onsagerCoeff}) indicates that the diagonal contributions
are in fact more significant.
Hence, although there are obvious qualitative differences in
the off-diagonal contributions to the diffusive flux in the
+HI and --HI simulations, the data suggest that it is important to include
the diagonal contributions to the flux when modeling stratification in nondilute
polymer mixtures, and HI modify the values of these coefficients.

\section{Conclusions}
\label{sec:conclusions}
We have investigated the microstructures of drying droplets containing mixtures
of short and long polymers using computer simulations. By employing models with
the same treatment of the polymers but different treatments of the solvent,
we have focused in particular on the role of hydrodynamic interactions (HI)
between polymers in setting the morphology of the dried supraparticle. In
qualitative agreement with prior studies of polymer mixtures in drying thin films,
we found that the polymers formed a core--shell morphology (with the short polymers
being enriched in the shell) in the simulations without HI, and this stratification
became more pronounced with increased evaporation speed. However, the morphology
remained homogeneous when HI were included.

We rationalized this behavior using a phenomenological multicomponent diffusion
model that connects the diffusive flux of the polymers to gradients in their
chemical potentials through effective Onsager coefficients. We measured these
coefficients directly in bulk polymer mixtures using nonequilibrium simulations,
finding that HI qualitatively altered off-diagonal coefficients such as
the one coupling the flux of long polymers to the chemical potential gradient
of the short polymers. However, we found that the diagonal coefficients
also played an important role in nondilute mixtures due to the relative magnitudes
of the chemical potential gradients. Indeed, the model predicted diffusive fluxes
consistent with the presence (or absence) of stratification in the drying
simulations when taking all coefficients into account in
combination with approximate expressions for the polymer chemical potentials.

Our simulations suggest several promising directions for improving models
for evaporation-induced stratification. First, although we were able to rationalize
the morphologies in our drying simulations through effective Onsager
coefficients, measuring these coefficients can be cumbersome. It will be important
to develop theoretical models or constitutive relations that robustly predict
Onsager coefficients for different types of polymers or mixtures. For example,
mixtures of polymers with chemical incompatibilities may give rise to
interesting morphologies, but their Onsager coefficients likely differ from those of the
polymers in good-solvent conditions that we studied \cite{RamirezHinestrosa:2020gd}.
Second, although we found that there were quantitative differences in the effective Onsager coefficients
when solvent backflow was accounted for in simulations without HI, these coefficients
were still qualitatively different from those in the simulations with HI. This discrepancy suggests
that it is essential to incorporate not only solvent backflow but also
hydrodynamic coupling between solutes in simulations and theoretical models of
stratification. Last, we focused in this work on the role of HI in the stratification of polymer
mixtures. It has been suggested that HI play a qualitatively similar role in the
stratification of drying colloidal mixtures; however, there may be quantitative
differences due to fundamental hydrodynamic differences between colloids and polymers.
A recent study attempted to address this question but used explicit-solvent and
implicit-solvent simulation models with interactions that were imperfectly matched \cite{Tang:2019jy},
making it difficult to unambiguously identify the role of HI.
A methodology similar to the one we used in this work should help to clarify this question, although
a different model for HI is likely required.

\section*{Author contributions}
\textbf{Michael P. Howard:} Conceptualization, Methodology, Software, Formal analysis, Writing, Visualization.
\textbf{Arash Nikoubashman:} Conceptualization, Methodology, Investigation, Resources, Writing, Funding acquisition.

\begin{acknowledgments}
We thank the organizers and participants of the CECAM workshop ``Applications of Diffusiophoresis in
Drying, Freezing, and Flowing Colloidal Suspensions'' for stimulating discussions and hospitality
during the time in which this work was initiated and Antonia Statt for providing helpful comments on this manuscript.
MPH acknowledges support from the Center for Materials for Water and Energy Systems,
an Energy Frontier Research Center funded by the U.S. Department of Energy,
Office of Science, Basic Energy Sciences under Award \#DE-SC0019272.
AN acknowledges funding through the German Research Foundation (DFG) under projects NI 1487/2-1 and GRK2516 (\#405552959).
Computing time was granted on the supercomputer Mogon at Johannes Gutenberg University Mainz
(www.hpc.uni-mainz.de).
\end{acknowledgments}

\section*{Data availability}
The data that support the findings of this study are available from the
authors upon reasonable request.

\bibliography{references}

\end{document}